\documentstyle[12pt]{article}

\begin{document}
\baselineskip=1.5\baselineskip
\renewcommand{\thefootnote}{\fnsymbol{footnote}}

\setcounter{equation}{0}
\setcounter{section}{0}
\renewcommand{\thesection}{\arabic{section}}
\renewcommand{\theequation}{\arabic{equation}}

\pagestyle{plain}
\begin{titlepage}

\hfill{ ANL-HEP-PR-94-02 ; SPhT/94-054}
\vfill
\begin{center}
{\large{\bf  {The Three Loop Equation of State of QED\\ }
{at High Temperature}}}\par
\end{center}
\vskip 1.2cm
\begin{center}
{Claudio Corian\`{o}$^{a,}$\footnote{email :
coriano@hep.anl.gov} \ and \ Rajesh R. Parwani$^{b,}
$\footnote{email : parwani@wasa.saclay.cea.fr}}
\end{center}
\vskip 0.8cm

\centerline{$^a$High Energy Physics Division}
\centerline{Argonne National Laboratory}
\centerline{9700 South Cass,Il 60439, USA.}
\vskip 0.5cm
\centerline{$^b$Service de Physique Th{\'e}orique, CE-Saclay}
\centerline{ 91191 Gif-sur-Yvette, France.}
\vskip 1.0cm
\centerline{PACS 12.20.Ds, 52.60.+h, 11.15.Bt, 12.38.Mh.}

\centerline{30 April 1994; Revised 30 August 1994. }

\vskip 1.0 cm
\centerline{\bf Abstract}

We present the three loop contribution (order $e^4$)
to the pressure of massless quantum electrodynamics at nonzero
temperature.
The calculation is performed within the imaginary time
formalism. Dimensional regularization is used to handle
the usual, intermediate stage, ultraviolet and infrared
singularities,
and also to prevent
overcounting of diagrams during resummation.\\

\end{titlepage}

\newpage

The equation of state (EOS) of relativistic quantum
electrodynamics (QED) at nonzero temperature ($T$)
and chemical potential ($\mu$) is of relevance in several
astrophysical contexts \cite{AP,FM,B,H}. It was obtained by
Akhiezer and Peletminskii to the third order ($e^3$) more than
three decades ago while the fourth order ($e^4$) contribution
at $T=0$, but nonzero $\mu$, was then obtained for QED and
quantum chromodynamics
(QCD) by Freedman and McLerran \cite{FM}, and Baluni \cite{B}.
However, to date, the analogous 3-loop calculation for $T\neq 0$
has,
to our knowledge, not been performed, presumably because of the
greater
technical difficulty in dealing with overlapping diagrams in the
presence
of Bose-Einstein and Fermi-Dirac statistical factors.

In this Letter we present the order $e^4$ contribution to the
pressure of a
QED plasma at $T\neq 0$ (and $\mu =0$), but for the case of
massless electrons.
This will also be the leading contribution at high
temperature for realistic QED with massive electrons.

Our motivation for this high order calculation is two-fold.
Firstly, for phenomenological applications
it is helpful to know how big corrections to the lower
order EOS can be. Secondly, it serves as a prototype to
illustrate techniques
which may be be used to perform similar calculations in QCD.
Recall that in QCD asymptotic freedom suggests \cite{CP} that at
high $T$ and/or density hadronic matter will transform into a
weakly interacting quark-gluon plasma, a novel state of matter
that is currently under intense study \cite{QM}. The EOS for
the high-temperature phase of QCD was determined by Kapusta
to third order ($g^3$) \cite{K} while Toimela \cite{T} has
extracted also the $ g^4 ln\,g$ piece. However, the
normalization of the logarithm in the last term requires
knowledge of the 3-loop contribution which is still lacking.

Returning to QED, let us introduce our notation and conventions.
We employ the imaginary time formalism whereby the energies take
on discrete Matsubara values, $q_0= i n \pi T $, $n$ being
an  even (odd) integer
for bosons (fermions).
Dimensional regularisation is be used to handle the ultraviolet
(UV) and infrared (IR) singularities which occur at intermediate
steps. The $D$ dimensional vector, $Q_\mu = (q_0,\vec{q})$ is
contracted with a Minkowski metric, $Q^2=q_0^2 - \vec{q}^2$.
In order to keep track of the
even (odd) Matsubara frequencies, we introduce the following
notation:

\begin{eqnarray}
\int [dq]\equiv T \sum_{q_0,even}\int {d^{D-1}q\over
(2 \pi)^{D-1}} \, , \nonumber
\end{eqnarray}

\begin{eqnarray}
\int\{dq\}\equiv T\sum_{q_0,odd}\int {d^{D-1}q\over
(2 \pi)^{D-1}} \, . \nonumber
\end{eqnarray}
The fermions are kept as four-component objects, $Tr(\gamma_{
\mu}\gamma_{\nu})=4 g_{\mu\nu}$, and for simplicity
we will work in the Feynman gauge so that the gauge propagator
is $g_{\mu\nu}/K^2$.
Renormalization via minimal subtraction ensures that the coupling
constant is gauge-fixing independent, and hence so will then be
our
final answer for the pressure \cite{B}.

Before discussing the 3-loop calculation, let us summarize the
lower order
results for the pressure of QED with $N$ massless Dirac fermions :

\begin{eqnarray}
P= P_0 +P_2 + P_3 +  \,\,\,O(e^4) \, ,
\label{pressure}
\end{eqnarray}
where
\begin{eqnarray}
 P_0 &=& {\pi^2\over 45}T^4 \left( 1 + {7\over 4}N\right) \, ,\\
\label{2b}
&& \nonumber \\
P_2 &=& -{5 e^2 T^4 N\over 288} \, ,\\
\label{2c}
&& \nonumber \\
P_3 &=& {e^3 T^4 \over 12 \pi}\left(N \over 3 \right)^{3/2} \, .
\label{2d}
\end{eqnarray}

The ideal gas contribution $P_0$ is determined by the one-loop
diagrams
in Fig.~$1$. The first correction, $P_2$, is given by Fig.~2.
The nonanalytic $(e^2)^{3/2}$ contribution, $P_3$, is a
consequence of Debye
screening \cite{BP}. In a perturbative  expansion using bare
propagators one discovers
power-like IR singularities in diagrams such as Fig.~3,
corresponding to
the $n=0$ Matsubara frequency of the photon propagator.
The diagrams in
Fig.~3 are singular because the electric polarization operator
behaves as
$\Pi_{00}(q_0,q\to 0)=m^2$, where $m$ is the electric screening
mass to
lowest order. Summing the infrared divergent pieces of the
diagrams in
Fig.~3 yields
\begin{eqnarray}
P_3 &=& {T\over 2}\sum_{p=2}^{\infty}\int {
d^3 q\over (2 \pi)^3} {(-1)^p\over p}
\left( {m^2\over q^2}\right)^p \label{3a} \\
&& \nonumber \\
\,\,\,\,\,\, &=& - {T\over 2}\int {d^3q\over (2 \pi)^3}
\left( \ln(1+{m^2\over q^2}) -{m^2\over q^2} \right).
\label{3b}
\end{eqnarray}

Although (\ref{3b}) is UV and IR finite and may be evaluated
directly to give
(\ref{2d}), it will be helpful to reconsider it using dimensional
continuation. Then \cite{P1} the second term in (\ref{3b})
vanishes and the first term gives
\begin{eqnarray}
{T\over 2} \  \Gamma \left( {1-D \over 2}\right) {\left(m^2 \over
4 \pi\right)}^{(D-1)/2} \, , \nonumber
\end{eqnarray}
which is $P_3$ as $D\to 4$.
Note that in dimensional regularization (DR) each single integral
of the series in (\ref{3a}) vanishes.
Neverthless the result in (\ref{3b}) is physically correct, and
mathematically nonzero, because the sum
in (\ref{3a}) must be done $before$ the integral.
This is obvious once one starts from an expansion of the
pressure in terms
in  terms of the full propagator \cite{AP,FM,B}.
Though the infinite sum must be performed first,
we may separate out any $finite$ number of terms from the sum,
and  for these we may use DR to deduce a zero contribution.
This explains why the second term in (\ref{3b}) may be dropped
in DR and we will exploit this fact further below.

Consider now the 3-loop diagrams. The order $e^4 N$
diagrams are shown in Fig.~4 and 5.
However, for massless fermions, the Ward identity $Z_1=Z_2$
implies
the mutual cancellation of the counterterm diagrams. Thus the sum
$G_1\,+\,G_2$ (Fig. 4) is UV finite.
After performing the spinor traces, some algebraic manipulation,
 and the use of scaling arguments such as in \cite{AE}, we obtain
\begin{eqnarray}
{G_1 + G_2 \over e^4 \ N \ T^{3 D-8} \ \mu^{8-2 D}}& =& { (D-2)
\over 6} \left(\ 2 \ (1-2^{(11-3D)}) \ H_1 \ + \ (20-3D) \
H_2 \ \right) \nonumber \\
&& \ + \ (D-2)^2 \ ( \ 2 \ H_3 \ - \ f_2 \ (f_1-b_1)^2 ) \, .
\nonumber
\end{eqnarray}
Here $\mu$ is the mass parameter of DR and $e$ is
the dimensionless, renormalized coupling. We have scaled all the
momenta by $1/T$ so that the integrals are dimensionless
(i.e. $T=1$ there) and are defined by
\begin{eqnarray}
b_n \equiv \int  {[d\,Q]\over (Q^2)^n} \, , \nonumber
\end{eqnarray}

\begin{eqnarray}
f_n \equiv  \int { \{d\,Q\} \over (Q^2)^n} \, , \nonumber
\end{eqnarray}

\begin{eqnarray}
H_1=\int {[d\,Q\,d\,P\,d\,K]\over K^2 Q^2 P^2 (K+Q+P)^2} \, ,
\nonumber
\end{eqnarray}

\begin{eqnarray}
H_2=\int {\{d\,K\,d\,R\,d\,S\}\over K^2 R^2 S^2 (K+R+S)^2} \, ,
\nonumber
\end{eqnarray}

\begin{eqnarray}
H_3=\int {\{d\,K\}[d\,Q\,d\,P]\, (P\cdot Q)\over K^2 P^2 Q^2
(K+Q)^2 (K+P)^2} \, . \label{h3}
\end{eqnarray}

The integral $H_1$ occurs in the  3-loop evaluation of the
pressure in $\phi^4$ theory \cite{Frenkel}, and  $H_2$, being
simply the
fermionic analog of $H_1$,  may be analysed in a similar manner.
The only
new integral left is $H_3$. Let us however first discuss the
order $e^4 N^2$ diagrams.
The UV singularity of $G_3$ (Fig.~6a) is cancelled by the
photon wave-function
renormalization through diagram, $X_1$ (Fig. 6b).
Diagram $G_3$ also has an IR singularity which
is precisely the first term of the series in eq.(\ref{3a}).
Since this term has already been considered there, it should be
subtracted
from $G_3$ to avoid overcounting.
However, as discussed earlier, this single term by itself
vanishes in DR, so double-counting is automatically avoided.
We have
\begin{eqnarray}
G_3 &=& {e^4 \ N^2 \over 4} \ T^{3 D-8} \ \mu^{8-2 D} \ 16 \ \left(
(D-4) \ b_2 \ f_1^2 + {(D-4)\over 4} \ H_2 + 4 \ H_4\right) \, ,
\label{8a}
\end{eqnarray}
and
\begin{eqnarray}
X_1=-(Z_3 -1) \ e^2 \ N \ (D-2) \ T^{3 D-8} \ \mu^{8-2 D} \ f_1 \
(2 b_1 - f_1) \
\left(T\over \mu\right)^{4-D} \; , \nonumber
\end{eqnarray}

where
\begin{eqnarray}
Z_3-1={e^2 \ N \over 6 \pi^2 (D-4)} \nonumber
\end{eqnarray}
to leading order. The factor
$\left({T\over \mu}\right)^{4-D}$ will generate a $e^4 \ln\,T/\mu$
term on expansion, but this will be reabsorbed later in the
definition of the
temperature dependent coupling $e(T)$. The new integral in
(\ref{8a}) is

\begin{eqnarray}
H_4=\int {[d\,Q]\{d\,K\,d\,R\} \ (K\cdot R)^2\over Q^4 K^2 R^2
(Q+K)^2 (Q+R)^2} \ . \nonumber
\end{eqnarray}

We now sketch our evaluation of $H_3$ (\ref{h3}). The
frequency sums are first rewritten in terms of contour-integrals as
in \cite{FM,B} but we
do not separate the $T=0$ and  $T\neq 0$ parts
before going to the ``phase-space''
representation. In this way we obtain

\begin{eqnarray}
H_3 = J_1+K_1 +L_1. \nonumber
\end{eqnarray}
The piece $L_1$ contains integrals which can be performed
analytically while
$K_1$ is an integral similar to $H_1$.
The difficulty lies in
\begin{eqnarray}
J_1\equiv \int {d^D\,K\, d^D\,Q\, d^D\,P\over (2 \pi)^{3(D-1)}}
\delta_+(K^2)\delta_+(P^2)\delta_+(Q^2)N_{k_0}n_{q_0}
(N_{p_0}+n_{p_0})\sum_{\sigma,\gamma=\pm1}(-\sigma)
S(\sigma,\gamma) \, ,
\nonumber
\end{eqnarray}
with
\begin{eqnarray}
S(\sigma,\gamma) \equiv {P\cdot Q\over K\cdot Q} \,
{1\over K\cdot Q + P\cdot (\sigma K + \gamma Q)} \, , \nonumber
\end{eqnarray}
and
\begin{eqnarray}
N_{k_0}={1\over e^{k_0}+1} \, . \nonumber
\end{eqnarray}
The integral $J_1$ is UV finite but has a collinear singularity.
We use the method of Sudakov decomposition, a technique which
 is well known
at $T=0$ (see for example \cite{S}) but appears to be novel
in this context,
to extract the  pole and the finite part of this integral.
The details are lengthy and will be presented elsewhere \cite{PC}.
Here we only state the final result,
\begin{eqnarray}
J_1={1\over 128 \pi^4}\left({r_1\over (D-4)} + c_1\right) \, ,
\nonumber
\end{eqnarray}
where
\begin{eqnarray}
r_1 &=& -0.7167667897 \pm 10^{-10} \nonumber \, , \\
c_1 &=& 3.936 \pm 5 \times 10^{-3}. \nonumber
\end{eqnarray}
The residue $r_1$ is  obtained as a finite two-dimensional
integral which we have
then evaluated numerically to high precision (relative
error $10^{-10}$), and the pole in $J_1$ above cancels with the other
poles contributing to the sum of $G_1 + G_2$ (again to the same
precision).
For $H_4$, the analysis is similar to $H_3$ but more tedious because of
the doubled
propagator $1/(Q^2)^2$.

Collecting all the pieces, we obtain
\begin{eqnarray}
{P_{4} \over e^4 T^4} &=& {(G_1 + G_2 +G_3 +X_1) \over e^4 T^4}
\nonumber \\
&& \nonumber \\
&= &{ N\over \pi^6}(0.4056) -  N^2
\left( {0.4667\over \pi^6} +{5\over 6\pi^2 \times 288}
\ln {T\over \mu}\right).
\label{banana}
\end{eqnarray}

The pressure up to and including order $e^4$ then follows from
eqs.(\ref{pressure}-\ref{2d}) and (\ref{banana}). It may be
rewritten in terms of the
(one-loop) renormalization group invariant coupling, at the
energy scale $T$, given by
\begin{eqnarray}
e^2(T)=e^2\left( 1+ {e^2 N \over 6 \pi^2} \ln {T\over \mu}\right)
\ . \nonumber
\end{eqnarray}
Defining $\alpha(T)={e^2(T)/ 4\pi}$ we finally arrive at
\begin{eqnarray}
P\over T^4 &=& {\pi^2\over 45} \ (1 +{7\over 4}N) \ - \
{5\pi^2\over 72} \ {\alpha(T)N\over \pi} \ + \ {2 \pi^2
\over 9 \sqrt{3} }
\left({\alpha(T) N\over \pi}\right)^ {3/2} \nonumber \\
&& \nonumber \\
&&\,+\left({0.658 \pm  0.006  \over N}
- 0.757  \pm 0.004 \right)\left({\alpha(T) N \over
\pi}\right)^2 + O\left(\alpha(T)^{5/2}\right). \nonumber \\
&& \label{papaya}
\end{eqnarray}

This (\ref{papaya}) is our expression for the three-loop pressure of
QED with $N$ electron flavours at high temperature; if $m_e$
is the (zero temperature) electron mass, we require
${m_e / T} \ll \alpha(T)^2$ so that mass corrections are subleading
to the terms displayed  in (\ref{papaya}).
Real world QED corresponds to $N=1$ and
in the regime where $\alpha(T) \ll 1$, the three
loop contribution is found to be indeed  a small correction.
However, one should note that since perturbative QED is not
asymptotically free, the effective coupling $\alpha(T)$
increases with temperature (albeit slowly) so that at
sufficiently high temperatures the fourth order contribution
becomes relevant.
Defering further discussion and potential applications of
(\ref{papaya}) to a later stage \cite{PC}, we mention that the
unknown $e^5$  contribution in eq.(\ref{papaya})
is a higher order analog of the
$e^3$ plasmon term and is also calculable \cite{P2}.

We conclude by summarizing the steps leading from the Feynman
diagrams to the result
(\ref{papaya}) : (i) algebraic reduction of the integrals,
(ii) evaluation of
frequency sums  by a contour-integral algorithm,
(iii) evaluation of
the final phase-space-like integrals, in particular using the
method of
Sudakov variables for UV finite integrals with a collinear
singularity,
and, (iv) use of dimensional regularization to
simplify the prevention of double counting during resummation.
We hope that the  methodology adopted here
opens the way for a similar, and long awaited,
calculation in QCD.\\

\smallskip
\smallskip
Acknowledgments: We thank  J.P. Blaizot, H. Contopanagos,
L.D. McLerran,\\
J-Y. Ollitrault, D.K. Sinclair, A.R. White and C. Zachos for
discussions.
We also thank P. Arnold and C. Zhai for pointing out errors in
our original results and for communicating with us their
calculation of the three-loop free-energy of Yang-Mills theory
\cite{AZ}.

\pagebreak
\noindent{\underline{\Large Figure Captions}}
\smallskip

\underline{Fig.1}:\\ Contribution to the ideal gas pressure.
The wavy line represents the photon propagator.

\underline{Fig.2}: \\ The two loop diagram.

\underline{Fig.3}:\\  Diagrams contributing to the $e^3$
plasmon term. The self-energy insertions are $\Pi_{00}(0,0)$
and the photon is static, $q_{0}=0$.

\underline{Fig.4}:\\
The order $e^4 N$ contributions : $G_1$ and $G_2$.

\underline{Fig.5}: \\
Ultraviolet counterterm diagrams for Fig.4.

\underline{Fig.6}:\\ Fig.6a is the $e^4 N^2$ contribution
($G_3$) while Fig.6b is the corresponding counterterm
diagram $X_1$.
\vfil

\pagebreak

\end {document}